\documentclass[preprint,pra,amsmath,amssymb]{revtex4}
\usepackage{graphicx}
\usepackage{dcolumn}
\usepackage{epic}%
\usepackage{eepic}%
\usepackage{bm}

\begin{document}

\preprint{APS/123-QED}

\title{The Conversion of nonlocal one-body operators into local ones: The Slater
  potential revisited}

\author{James P. Finley}

\affiliation{
Department of Physical Sciences,
Eastern New~Mexico University,
Station \#33, Portales, NM 88130}
\email{james.finley@enmu.edu}

\affiliation{Department of Applied Chemistry, Graduate School of Engineering,
The University of Tokyo, Tokyo, Japan 113-8656}

\date{\today}

\begin{abstract}

One-particle Schr\"odinger equations are considered, e.g., the Hartree--Fock
equations, that contain a nonlocal operator, e.g., the Hartree--Fock exchange
operator, where this operator depends on the one-particle density-matrix of a
determinantal state.  One-body nonlocal operators of this type are converted into
approximate local potentials that depend on the kernel of the nonlocal operator
and, also, the one-particle density matrix that, as mentioned above, the nonlocal
operator also depends on. When the non-local operator is the exchange operator,
the method yields the Slater potential.

\end{abstract}

\maketitle

\newcommand{\sizet}{\tiny}
\newcommand{\labelz}[1]{\label{#1}}
\newcommand{\ee}[1]{
\begin{eqnarray} #1
\end{eqnarray}}
\newcommand{\fc}{\frac} 
\newcommand{\lt}{\left} 
\newcommand{\rt}{\right} 
\newcommand{\mbt}[1]{\mbox{\tiny #1}} 
\newcommand{\nn}{\nonumber} 

\newcommand{\mr}{\mathbf{r}}         
\newcommand{\mx}{\mathbf{x}}         
\newcommand{\roo}{{\rho_1}}          
\newcommand{\kp}{\kappa}             
\newcommand{\dt}{\delta}             
\newcommand{\hF}{\hat{F}}            
\newcommand{\hw}{\hat{w}}            
\newcommand{\gm}{{\gamma}}           
\newcommand{\si}{{\sigma}}           
\newcommand{\sip}{{\sigma^{\prime}}} 
\newcommand{\sipp}{{\sigma^{\prime\prime}}} 
\newcommand{\sit}{{\tilde{\sigma}}}  
\newcommand{\om}{{\omega}}           
\newcommand{\roos}{{\rho_{1\sigma}}} 
\newcommand{\roosp}{{\rho_{1\sigma^\prime}}} 
\newcommand{\pr}{{\prime}}           

\section{Introduction}

There is a considerable interest in the conversion of nonlocal, one-body operators
into local, multiplicative operators, or potentials.  For example, using a local,
multiplicative operator to replace a nonlocal one can lead to significant
computational savings when solving the Hartree--Fock equations, or other similar
one-body, coupled equations.  In addition, the Kohn-Sham approach of density
functional theory (DFT)
\cite{Kohn:65,Dreizler:90,Parr:89,Springborg:97,Ellis:95,Gross:94,Seminario:95}
requires the noninteracting state to be obtained from a local potential, since
this formalism invokes the Hohenberg-Kohn theorem \cite{Hohenberg:64a}, even
though, according to the adiabatic connection, the exact exchange energy is the
same one-particle density-matrix functional as the one from Hartree--Fock theory
\cite{Becke:93b,Harris:74,Gunnarsson:76,Langreth:77,Harris:84}, except that the
one-particle functions are Kohn--Sham orbitals.

Other possible applications of nonlocal to local conversions is for one-particle
density-matrix approaches that employ a nonlocal operator
\cite{Gilbert:75,Berrondo:75,Donnelly:78,Levy:79,Valone:80,Ludena:87}. For
example, in the reference-state one-particle density-matrix formalism
\cite{Finley:bdmt,Finley:bdft}, the conversion of the nonlocal
exchange-correlation potential into a local operator leads naturally to a density
functional theory, since the one-particle density matrix of the determinantal
reference state obtained from a local potential is a unique functional of its
electron density, according to the Hohenberg-Kohn theorem.

A classical example of the conversion of a nonlocal operators into a local one is
Slater's local exchange-operator, derived from a uniform electron gas
\cite{Slater:51,Harbola:93}. A very similar approximate functional has also been
derived by Gasper \cite{Gasper:54}, where this one only differs by a constant from
the one derived by Slater.  The $X\alpha$ approach \cite{Slater:72} includes a
semiempirical parameter that yields Slater's original prescription with $\alpha =
1$, and the one by Gasper for $\alpha = \frac23$. Gasper's exchange operator is
also used within the local density approximation (LDA) of DFT
\cite{Kohn:65,Dreizler:90,Parr:89}. When the Gasper potential is combined with
Becke's exchange correction \cite{Becke:88}, derived in 1988, a very accurate
approximation is obtained, and this two component exchange potential is used
within the very successful DFT hybrid approach known as the B3LYP method
\cite{Becke:93,Stephens:94}.

The optimized potential method (OPM)
\cite{Fiolhais:03,Sharp:53,Talman:76,Li:93,Shaginyan,Goorling:94,Grabo:00} is an
approach to convert a nonlocal operator into a local potential.  This method,
unfortunately, leads to rather complicated functionals that depend on the
individual orbitals.  Below we introduce an approach that converts one-body
nonlocal operators into approximate local potentials that depend on the kernel of
the nonlocal operator and, also, the one-particle density matrix that the nonlocal
operator also depends on. When the non-local operator is the exchange operator,
the method yields the Slater potential \cite{Slater:51,Harbola:93,Hirata:01},
which also appears as the leading term from the Krieger--Li--Iafrate (KLI)
approximation of the OPM \cite{Fiolhais:03,Li:92,Li:93,Hirata:01}.

\section{The closed-shell restricted spin-orbital formalism \labelz{7278} } 

Consider a set of spatially restricted spin-orbitals \cite{Szabo:82}:
\begin{eqnarray} 
\psi_{i\sigma}(\mr_1,\om) = \chi_i(\mr_1)\sigma(\om_1), \; \sigma = \alpha,\beta, 
\end{eqnarray}
where the spin and spatial coordinates are given by $\mr_1$ and $\om_1$,
respectively.  Consider also the following spinless, one-particle Schr\"odinger
equation:
\begin{eqnarray} \labelz{3969}
\hF_\roo \chi_i(\mr_1) = \epsilon_i \chi_i(\mr_1),
\end{eqnarray}
where this Hermitian one-body operator is given by
\begin{eqnarray} \labelz{8422}
\hF_\roo = -\frac12 \nabla^2 + v_\roo + \hw_\roo,
\end{eqnarray}
and $v_\roo$ and $\hw_\roo$ are local and nonlocal operators, respectively; these
operators depend on the spinless one-particle density matrix $\roo$ of a closed
shell single-determinantal state, say $|\Phi\rangle$; that is, we have
\cite{Parr:89,McWeeny:60,Finley:bdft}
\begin{eqnarray} \labelz{8373}
\roo(\mr_1,\mr_2) = 2 \sum_w \chi_w(\mr_1) \chi_w^*(\mr_2),
\end{eqnarray}
where the same spatial-orbitals appearing in this summation are doubly occupied
within $|\Phi\rangle$; henceforth, $w$, $x$ denote these occupied orbitals from
$|\Phi\rangle$; $r$, $s$ denote the unoccupied orbitals.

Now consider the possibility of replacing the nonlocal operator $\hw_\roo$ by a
local one, say~$z_\roo$; so, we have
\begin{eqnarray} \labelz{6723}
z_\roo(\mr_1)\chi_x(\mr_1)
=
\int d\mr_2 \; w_\roo(\mr_1,\mr_2)\chi_x(\mr_2),
\end{eqnarray}
where $w_\roo(\mr_1,\mr_2)$ is the kernel of $\hw_\roo$. 

Our interest here is in the one-particle density-matrix $\roo$ that arises from
solving Eq.~(\ref{3969}). Therefore, we only consider the operator $\hw_\roo$
acting upon an occupied orbital, $\hw_\roo\chi_x$ in the above equation. In other
words, $\roo$ does not depend on $\hw_\roo\chi_r$, where $\chi_r$ is an excited
orbital.  We do, however, require $\hw_\roo$ to be Hermitian.

Multiplying the previous equation by $\chi_x^*(\mr_3)$ and summing over the
orbital indices gives
\begin{eqnarray} \labelz{7892}
z_\roo(\mr_1)\roo(\mr_1,\mr_3) =
\int d\mr_2 \; w_\roo(\mr_1,\mr_2)\roo(\mr_2,\mr_3).
\end{eqnarray}
Setting $\mr_3= \mr_1$ yields the desired result:
\begin{eqnarray} \labelz{8892}
z_\roo(\mr_1) \approx
\rho^{-1}(\mr_1) \int d\mr_2 \; w_\roo(\mr_1,\mr_2)\roo(\mr_2,\mr_1),
\end{eqnarray}
where $\rho(\mr)$ is the electron density, $\roo(\mr,\mr)$.  For example, the
exact exchange operator from Hartree--Fock theory yields the Slater potential
\cite{Slater:51,Harbola:93,Hirata:01},
\begin{eqnarray} \labelz{8262}
v_{\roo}^{\mathrm{x}}(\mr_1)
\approx
-\frac12 \rho^{-1}(\mr_1) \int d\mr_2 \; r_{12}^{-1}|\roo(\mr_1,\mr_2)|^2,
\end{eqnarray}
where the kernel of the exchange operator is
$-\frac12r_{12}^{-1}\roo(\mr_1,\mr_2)$.

In the two previous expressions above we have changed the equality to an
approximation, since these are apparently not identities, where this conclusion
arises, in part, since, in our derivation of Eq.~(\ref{7892}) we have summed over
all occupied orbitals. Now if Eq.~(\ref{6723}) is valid, of course,
Eqs.~(\ref{7892}) and (\ref{8892}) must be satisfied, but, not vice versa. So if
we define $z_\roo$ by Eq.~(\ref{8892}) we will probably not satisfy
Eq.~(\ref{6723}). For example, consider a simple case of only two occupied
orbitals, say $\chi_w$ and $\chi_x$, where instead of Eq.~(\ref{6723}) being
satisfied, we have the following relations:
\begin{subequations}
\labelz{4439}
\begin{eqnarray}
z_\roo(\mr_1)\chi_x(\mr_1)
&=&
\int d\mr_2 \; w_\roo(\mr_1,\mr_2)\chi_x(\mr_2) + \chi_y^*(\mr_1)\phi(\mr_1),
\\
z_\roo(\mr_1)\chi_y(\mr_1)
&=&
\int d\mr_2 \; w_\roo(\mr_1,\mr_2)\chi_y(\mr_2) - \chi_x^*(\mr_1)\phi(\mr_1),
\end{eqnarray}
\end{subequations}
where $\phi(\mr_3)$ is an arbitrary function.  Multiplying the first equation by
$\chi_x(\mr_3)^*$ and the second one by $\chi_y(\mr_3)^*$ removes the last terms
from both equations when the two equations are added, yielding Eq.~(\ref{7892}).
 
As in any operator, our operators are completely defined by their matrix
elements. If the following identities are satisfied, then the local and nonlocal
operators are equivalent:
\begin{subequations}
\labelz{7393}
\begin{eqnarray} \labelz{7341}
\int d\mr_1 \; \chi_y(\mr_1) z_\roo(\mr_1)\chi_x(\mr_1) 
&=&
\int d\mr_1\int d\mr_2 \; \chi_y(\mr_1) w_\roo(\mr_1,\mr_2)\chi_x(\mr_2),
\\ \labelz{7342}
\int d\mr_1 \; \chi_r(\mr_1) z_\roo(\mr_1)\chi_x(\mr_1) 
&=&
\int d\mr_1\int d\mr_2 \; \chi_r(\mr_1) w_\roo(\mr_1,\mr_2)\chi_x(\mr_2),
\\ \labelz{7343}
\int d\mr_1 \; \chi_r(\mr_1) z_\roo(\mr_1)\chi_s(\mr_1) 
&=&
\int d\mr_1\int d\mr_2 \; \chi_r(\mr_1) w_\roo(\mr_1,\mr_2)\chi_s(\mr_2),
\end{eqnarray}
\end{subequations}
and we can replace the one-body operator, given by Eq.~(\ref{8422}), by the following:
\begin{eqnarray} \labelz{8428}
\hF_\roo = -\frac12 \nabla^2 + v_\roo + z_\roo.
\end{eqnarray}
However, for our purposes, we do not need all three relations given by
Eqs.~(\ref{7393}) to be satisfied. In particular, if Eq.~(\ref{7343}) is not
satisfied, Eq.~(\ref{3969}) is still satisfied for the same occupied orbitals and,
as mentioned previously, the one-particle density-matrix $\roo$ is not
changed. Furthermore, if Eq.~(\ref{7341}) is not satisfied, we will get different
occupied orbitals but they will differ only by a unitary transformation as long as
Eq.~(\ref{7342}) remains valid, and, again, the one-particle density-matrix $\roo$
is not changed. Hence, we only need Eq.~(\ref{7342}) to be a reasonable
approximation.

\section{The open-shell unrestricted spin-orbital formalism \labelz{6383} } 

We now generalize the previous derivation to the case where the determinantal
state is composed of orbitals that are spatially unrestricted;
that is, we have \cite{Szabo:82}
\begin{eqnarray} \labelz{9282}
\psi_{i\sigma}(\mx_1) = \chi_{i\sigma}(\mr_1)\sigma(\om), \; \sigma =
\alpha,\beta,
\end{eqnarray}
where the spatial $\mr_1$ and spin coordinates $\om_1$ are denoted collectively by
$\mx_1$; furthermore, we have a different one-body operator for each
spin-function:
\begin{eqnarray} 
\hF_\gm^\si \chi_{i\si}(\mr_1) = \epsilon_{i\si} \chi_{i\si}(\mr_1),
\end{eqnarray}
where the Hermitian one-body operators are given by
\begin{eqnarray} 
\hF_\gm^\si = -\frac12 \nabla^2 + v_\gm^\si + \hw_\gm^\si,
\end{eqnarray}
and $v_\gm^\si$ and $\hw_\gm^\si$ are local and nonlocal operators, respectively;
these operators depend on the one-particle density matrix $\gamma$ of the
single-determinantal state $|\Phi\rangle$; that is, we have
\begin{eqnarray}
\gm(\mx_1,\mx_2) =  \gm(\mr_1,\om_1;\mx_2,\om_1) \dt_{\om_1\om_2},
\end{eqnarray}
where
\begin{eqnarray}
\gm(\mr_1,\om;\mx_2,\om) = \sum_\si |\si(\om)|^2 \roos(\mr_1,\mr_2), 
\end{eqnarray}
and for $N_\si$ occupied $\si$ orbital, the spin-components of $\gm$ are given by
\begin{eqnarray}\labelz{9283}
\roos(\mr_1,\mr_2) = \sum_w^{N_\si} \chi_{w\si}(\mr_1) \chi_{w\si}^*(\mr_2).
\end{eqnarray}

Our one-body operators $\hF_\gm^\si$, and potential, $v_\gm^\si$ and
$\hw_\gm^\si$, depend upon $\gm$. However, $\gm$ is an explicit functional of its
components, $\rho_{1\alpha}$ and $\rho_{1\beta}$; furthermore, we require
$\hF_\gm^\si$ not to depend on the spin variable $\om$, since this dependence is
easily removed. So, we could, instead write, for example,
$\hF_{\rho_{1\alpha},\rho_{1\beta}}^\si$. However, for a less cluttered notation,
we will continue to indicate a $\gm$ dependence. This is also not inaccurate,
since if we know $\gm$, we also know $\rho_{1\alpha}$ and $\rho_{1\beta}$.

Now, as in Eq.~(\ref{6723}), consider the possibility of replacing the nonlocal
operator $\hw_\gm^\si$ by a local one, say~$z_\gm^\si$:
\begin{eqnarray} 
z_\gm^\si(\mr_1)\chi_{x\si}(\mr_1)
=
\int d\mr_2 \; w_\gm^\si(\mr_1,\mr_2)\chi_{x\si}(\mr_2),
\end{eqnarray}
where $w_\gm^\si(\mr_1,\mr_2)$ is the kernel of $\hw_\gm^\si$.  Multiplying
this equation by $\chi_{x\si}^*(\mr_3)$ and summing over the orbital indices
gives
\begin{eqnarray} 
z_\gm^\si(\mr_1)\roos(\mr_1,\mr_3) =
\int d\mr_2 \; w_\gm^\si(\mr_1,\mr_2)\roos(\mr_2,\mr_3).
\end{eqnarray}
Setting $\mr_3= \mr_1$ yields the desired result:
\begin{eqnarray} \labelz{8638}
z_\gm^\si(\mr_1) \approx
\rho_{\sigma}^{-1}(\mr_1) \int d\mr_2 \; w_\gm^\si(\mr_1,\mr_2)\roos(\mr_2,\mr_1),
\end{eqnarray}
where $\rho_\si(\mr)$ is the $\si$-component of the electron density, given by
$\roos(\mr,\mr)$.  For example, the exact exchange operator from Hartree--Fock
theory yields the Slater potential \cite{Slater:51,Harbola:93,Hirata:01},
\begin{eqnarray} \labelz{9278}
v_{\roos}^{\mathrm{x}}(\mr_1)
\approx
- \rho_{\sigma}^{-1}(\mr_1) \int d\mr_2 \; r_{12}^{-1}|\roos(\mr_1,\mr_2)|^2,
\end{eqnarray}
where the kernel of the exchange operator is
$-r_{12}^{-1}\roos(\mr_1,\mr_2)$.

\appendix 

\section{The Slater Potential} \labelz{4282}

The Slater potential is a local approximation of the exchange potential that was
suggested by Slater and used to obtain the $X\alpha$ approach with ($\alpha = 1$).
While it is well known in the literature that Eq.~(\ref{9278}) is the Slater
potential, we are not familiar with a derivation demonstrating that is the
case. Therefore, we have decided to present one here. In appendix~\ref{7652} we
present a derivation of the X$\alpha$ exchange potential.

Using the notation by Slater \cite{Slater:51}, but in atomic units, Slater's potential
has the following form:
\begin{eqnarray} \labelz{8372}
v_{s}(\mx_1)
&=& 
-
\fc{
\sum_j^N
\sum_k^N
\int 
u_{j}^*(\mx_1) 
u_{k}^*(\mx_2)
u_{k}(\mx_1) 
u_{j}(\mx_2) 
r_{12}^{-1}
\; d\mx_2 
}{\sum_j^n u_{j}^*(\mx_1) u_{j}(\mx_1) },
\end{eqnarray}
where the sums are over the $N$ occupied spin orbitals from the set
$\{u_{j}\}$. Using our spin-orbitals, given by Eq.~(\ref{9282}), we have
\begin{eqnarray} \labelz{8375}
v_{s}(\mx_1)
&=& 
-
\fc{
\sum_{\si\sip}
\sum_x^{N_{\si}} 
\sum_y^{N_{\sip}} 
\int 
\psi_{x\si}^*(\mx_1) 
\psi_{y\sip}^*(\mx_2)
\psi_{y\sip}(\mx_1) 
\psi_{x\si}(\mx_2) 
r_{12}^{-1}
\; d\mx_2 
}
{\sum_{\si}
\sum_x^{N_\si} \psi_{x\si}^*(\mx_1) \psi_{x\si}(\mx_1)},
\end{eqnarray}
where the sums are over the $N_\alpha$ and $N_\beta$ occupied spin orbitals, and
from Eq.~(\ref{9282}), we have
\begin{eqnarray} 
v_{s}(\mx_1) = \hspace{79ex}
\\ \nn \mbox{}
-\fc{
\mbox{\footnotesize
$\sum_x^{N_{\si}} 
\sum_y^{N_{\sip}} 
\sum_{\si\sip}$}
\int 
\chi_{x\si}^*(\mr_1)  
\chi_{y\sip}^*(\mr_2) 
\chi_{y\sip}(\mr_1)
\chi_{x\si}(\mr_2)
r_{12}^{-1}
\; d\mr_2 
\si^*(\om_1)\sip(\om_1)
\sum_{\om_2}
\sip^*(\om_2)\si(\om_2)
}
{
\sum_x^{N_\si} 
\sum_{\si}
\chi_{x\si}^*(\mr_1) 
\chi_{x\si}(\mr_1)   
|\si(\om_1)|^2}.
\end{eqnarray}
Using the following identity:
\begin{eqnarray} 
\si^*(\om_1)\sip(\om_1) = \dt_{\si\sip}|\si(\om_1)|^2,
\end{eqnarray}
gives
\begin{eqnarray} 
v_{s}(\mx_1)
&=& 
-
\fc{
\sum_x^{N_{\si}} 
\sum_y^{N_{\si}} 
\sum_{\si}|\si(\om_1)|^2
\int 
\chi_{x\si}^*(\mr_1)  
\chi_{y\si}^*(\mr_2) 
\chi_{y\si}(\mr_1)
\chi_{x\si}(\mr_2)
r_{12}^{-1}
\; d\mr_2 
}
{
\sum_{\si}|\si(\om_1)|^2
\sum_x^{N_\si} 
\chi_{x\si}^*(\mr_1) 
\chi_{x\si}(\mr_1)   
},
\end{eqnarray}
which can be rearranged,
\begin{eqnarray} 
v_{s}(\mx_1)
&=& 
-
\fc{
\sum_{\si}|\si(\om_1)|^2
\int 
\lt(
\sum_y^{N_{\si}} 
\chi_{y\si}(\mr_1)
\chi_{y\si}^*(\mr_2) 
\rt)\lt(
\sum_x^{N_{\si}} 
\chi_{x\si}(\mr_2)
\chi_{x\si}^*(\mr_1)  
\rt)
r_{12}^{-1}
\; d\mr_2 
}
{
\sum_{\si}
|\si(\om_1)|^2
\lt(
\sum_x^{N_\si} 
\chi_{x\si}(\mr_1)   
\chi_{x\si}^*(\mr_1) 
\rt)
},
\end{eqnarray}
and by using Eq.~(\ref{9283}), we have
\begin{eqnarray} 
v_{s}(\mx_1)
&=& 
-
\fc{\sum_{\si}|\si(\om_1)|^2
\int 
|\roos(\mr_1,\mr_2)|^2
r_{12}^{-1}
\; d\mr_2 }
{
\sum_{\si}
|\si(\om_1)|^2 \rho_{\sigma}(\mr_1)}.
\end{eqnarray}
Furthermore, the following identity is readily verified:
\begin{eqnarray} 
v_{s}(\mr_1,\om_1) \chi_{i\si}(\mr_1) \si(\om_1)
&=& v_{\roos}^{\mathrm{x}}(\mr_1) \chi_{i\si}(\mr_1) \si(\om_1)
\end{eqnarray}
indicating that the Slater potential $v_{s}$, given by Eq.~(\ref{8372}) or
(\ref{8375}), is equivalent to our potential, $v_{\roos}^{\mathrm{x}}$, given by
Eq.~(\ref{9278}).

\section{The X$\alpha$ exchange potential} \labelz{7652}

It is well known that the spinless, one-particle density matrix
$\roo(\mr_1,\mr_2)$ of a closed-shell unified electron-gas depends on its density
$\rho$, a constant.  A generalized expression for $\roo$ is obtained by replacing
the constant density with a non-constant one, say $\rho(\mr)$ \cite{Parr:89}:
\begin{eqnarray}
\labelz{8529}
\roo(\mbox{$\mathbf{r_1}$},\mbox{$\mathbf{r_2}$})
&=&
\roo(\mr,s) =
3\rho(\mr)
\left( 
\frac{\sin{[k_f(\mr)s]}
- k_f(\mr)s
\cos{[k_f(\mr)s]}}{[k_f(\mr)s]^3}
\right),
\end{eqnarray}
where 
\begin{eqnarray}
\labelz{5830}
k_f(\mr) &=&  \sqrt[3]{3\pi^2\rho(\mr)},
\end{eqnarray}
and a change of coordinates is employed:
\begin{eqnarray}
\mr&=& \frac12 (\mathbf{r_1} + \mathbf{r_2}),\\
\mathbf{s}&=&\mr_1-\mr_2, \\
s&=&|\mathbf{s}| = r_{12} = \lt[(x_2-x_1)^2 + (y_2-y_1)^2 + (z_2-z_1)^2 \rt]^{1/2}.
\end{eqnarray}
Using this expression for $\roo$, the exchange-energy functional for a closed
shell system, given by
\begin{eqnarray}
E_{\mathrm{x}}[\roo] = 
-\frac14 \int  \int d\mr_1d\mr_2 \; r_{12}^{-1}|\roo(\mr_1,\mr_2)|^2,
\end{eqnarray}
leads to the well known Dirac exchange-energy density-functional
\begin{eqnarray}
E_{\mathrm{x}}[\rho] \approx -\frac{3}{4}\lt(\frac{3}{\pi}\rt)^{1/3}\int
\rho^{4/3}(\mathbf{r}) \,d\mathbf{r},
\end{eqnarray}
where in the derivation by Parr and Yang \cite{Parr:89}, the integration is
expressed using the $\mr$ and $\mathbf{s}$ coordinates and the integration is
carried out over $\mathbf{s}$. Analogous expressions for unrestricted orbitals and
open shell systems are also readily derived \cite{Parr:89,Gross:94}.

In the Kohn-Sham LDA \cite{Kohn:65,Parr:89,Gross:94}, the functional derivative of
the above exchange functional gives the following local exchange potential:
\begin{eqnarray}
v_{\mathrm{x}}^{\rho}(\mr)=-\lt(\frac{3}{\pi}\rho(\mathbf{r})\rt)^{1/3},
\end{eqnarray}
in agreement with Gasper's potential \cite{Gasper:54}. However, it well known
that this functional differs from the one obtained by Slater \cite{Slater:51} by a
factor of $\frac{2}{3}$, where $v_{\mathrm{x}}^{\rho}(\mr)$ is obtained directly
from the Hartree-Fock, nonlocal, exchange-operator and, in addition, Slater's
approach uses a sort of averaging over the occupied orbital states.

We now demonstrate that the X$\alpha$ exchange potential, with original
prescription of $\alpha = 1$, arises in our approach when considering a uniform
electron gas, where we only consider the closed-shell spin restricted formalism,
and the derivation in analogous to the one by Parr and Yang mentioned above. In
our derivation, we also use Eq.~(\ref{8529}) except that, at least for the moment,
we leave the density as a constant: \ee{\roo(\mr_1,\mr_2) = 3 \rho
\fc{1}{k_f^3s^3} \lt( \sin(k_fs) - k_fs \cos(k_fs) \rt).}  Substituting this
expression into Eq.~(\ref{8262}) and using Cartesian coordinates gives
\ee{
v_{\roo}^{\mathrm{x}}(\mr_1) = -\frac92 \rho \int\int\int dx_2\,dy_2\,dz_2\, 
\fc{1}{s^7k_f^6}
\lt( \sin(k_fs) - k_fs \cos(k_fs) \rt)^2.}
Now as far as the integral is concerned, $x_1$, $y_1$, are $z_1$ are
constants. Therefore, by making the following substitution:
\ee{s_x = (x_2 - x_1)} 
as well as analogous ones for $s_y$ and $s_z$, where, for example ($ds_x = dx_2$),
we get
\ee{
v_{\roo}^{\mathrm{x}}(\mr_1) = -\frac92 \rho \int \int \int ds_x\,ds_y\,ds_z\, \fc{1}{s^7k_f^6}
\lt( \sin(k_fs) - k_fs \cos(k_fs) \rt)^2,}
which in spherical coordinates, can be written as
\ee{ 
v_{\roo}^{\mathrm{x}}(\mr_1) = -18 \pi \fc{\rho}{k_f}\int ds \,\fc{1}{s^5k_f^5}
\lt( \sin(k_fs) - k_fs \cos(k_fs) \rt)^2.
}
Making the following substituting $t = k_fs$, where $ds = dt/k_f$, we have
\ee{
v_{\roo}^{\mathrm{x}}(\mr_1) = -18 \pi \fc{\rho}{k_f^2}\int_0^\infty dt \,\fc{1}{t^5}
\lt( \sin t - t \cos t \rt)^2.}
and since the integral is $\frac14$ \cite{Parr:89}, we get
\ee{
v_{\roo}^{\mathrm{x}}(\mr_1) = v_{\rho}^{\mathrm{x}} = -\frac92 \pi \fc{\rho}{k_f^2},}
where we have introduced a different notation for the exchange potential, since it
is simply a constant that depends on $\rho$. Using Eq.~(\ref{5830}), we obtain the
following expression after some algebra and by replacing $\rho$ by a non-constant
density, $\rho(\mr)$:
\ee{ 
v_{\rho}^{\mathrm{x}}(\mr) = -\frac{3}{2} \lt(\fc{3}{\pi}\rho(\mr)\rt)^{\mbt{$1/3$}}. \;
}
This expression is in agreement with Slater's original prescription of $\alpha
= 1$.

\bibliography{jfinley2}
\end{document}